\documentclass[preprint,preprintnumbers,amsmath,amssymb,nofootinbib]{revtex4}
\usepackage{graphicx,color}
\usepackage{amsmath,amssymb}
\usepackage{bm}

\newcommand\nn{\nonumber \\}

\newcommand{\mE}{\mathcal E}

\newcommand{\lk}{\left(}
\newcommand{\rk}{\right)}
\newcommand{\ltk}{\left\{}
\newcommand{\rtk}{\right\}}
\newcommand{\ldk}{\left[}

\newcommand{\rdk}{\right]}
\newcommand\beq{ \begin{eqnarray} }
\newcommand\eeq{ \end{eqnarray} }

\begin{document}

%\preprint{}
\title{Quasiparticle properties of a single $\alpha$ particle 
in cold neutron matter}
\author{Eiji Nakano}
\email{e.nakano@kochi-u.ac.jp}
\affiliation{Department of Mathematics and Physics, Kochi University, Kochi 780-8520, Japan}
\affiliation{Institut f\"{u}r Theoretische Physik, Goethe Universit\"{a}t Frankfurt, 
D-60438 Frankfurt am Main, Germany}
\author{Kei Iida}
\email{iida@kochi-u.ac.jp}
\affiliation{Department of Mathematics and Physics, Kochi University, Kochi 780-8520, Japan}
\author{Wataru Horiuchi}
\email{whoriuchi@nucl.sci.hokudai.ac.jp}
\affiliation{Department of Physics, Hokkaido University, Sapporo 060-0810, Japan}
\date{\today}

\begin{abstract}
 Light clusters such as $\alpha$ particles and deuterons are predicted to occur in
hot nuclear matter as encountered in intermediate-energy heavy-ion collisions and 
protoneutron stars.  To examine the in-medium properties of such light clusters,
we consider a much simplified system in which like an impurity, a single $\alpha$  
particle is embedded in a zero-temperature, dilute gas of non-interacting neutrons.
By adopting a non-selfconsistent ladder approximation for the effective 
interaction between the impurity and the gas, which is often used for analyses
of Fermi polarons in a gas of ultracold atoms,  we calculate the quasiparticle
properties of the impurity, i.e., the energy shift, effective mass, quasiparticle
residue, and damping rate.
\end{abstract} 

%\pacs{}
%\keywords{}

\maketitle

%\tableofcontents
\section{Introduction}
To understand the properties of hot nuclear matter as encountered in supernova 
cores is essential in describing various phenomena such as neutrino bursts, 
nucleosynthesis, and formation of neutron stars \cite{Bethe}.  Remarkably, this matter 
can be opaque even to neutrinos, which in turn play a role in carrying a released 
gravitational energy of order 10$^{53}$ erg during stellar collapse by diffusing 
out of supernova cores and at the same time in depositing a sufficient energy onto 
the material to cause a supernova explosion at which the total kinetic energy is 
of order 10$^{51}$ erg.  Here it is significant to note that at finite temperature,
a nonzero number of light clusters such as $\alpha$ particles, deuterons, tritons, and
$^{3}$He nuclei appear even in chemical equilibrium.  These clusters, if sufficiently
present in supernova cores, can play a role in scattering or absorbing the outgoing 
neutrinos.  For example, even an $^{56}$Fe nucleus, one of the most stable nuclear 
configurations in vacuum, can decompose into thirteen $\alpha$ particles and four 
neutrons by absorbing a gamma-ray photon of energy in excess of $-Q=124$ MeV.  
If an $^{56}$Fe nucleus is simply assumed to be a primary 
component of matter at typical conditions where the neutron fugacity is of 
order 0.1--1, 
according to Saha's arguments, the mass fraction of $\alpha$ particles
is dominated by a factor of $e^{Q/13k_B T}$, with the temperature $T$.  
In estimating such a fraction, however, no in-medium modification of nuclear masses 
except the Coulomb corrections is normally considered.  
To deal with the in-medium modification, we will focus on a polaron
picture, namely, a light cluster dressed by excitations in the medium.

%Theoretical investigations of hot nuclear matter are liquid-drop models,
%TF models, and HF models, virial or excluded-volume......
The earliest theoretical investigations of hot nuclear matter are based on 
liquid-drop models \cite{Lamb}.  The key ingredient of these models is mass formula for
neutron-rich nuclei. Typically, in the presence of trapped electron neutrinos, 
nuclei in such matter (so-called supernova matter) are not extremely neutron-rich 
but too neutron-rich for their masses to be measured, which requires an extrapolation 
from empirical mass data.  In the presence of internuclear Coulomb coupling, 
the Wigner-Seitz approximation for a lattice of nuclei embedded in a 
neutralizing background of electrons is often utilized.
This approximation is known to give a good estimate of the lattice energy.
Concerning the mass distribution, furthermore, a single species 
approximation in which only the nuclide that minimizes the system energy 
at fixed baryon density and lepton fraction occurs is often adopted for simplicity.  
This is good at sufficiently low temperatures.  For better estimates of the 
mass distribution at temperatures relevant for supernova cores, the presence 
of $\alpha$ particles etc.\ has been allowed for or even the nuclear statistical 
equilibrium has been imposed in some cases \cite{Buyukcizmeci}.  
The most important upgrade has been made on how to calculate the nuclear
mass itself as an extrapolation from the empirical data.  The nucleon density
profile in a Wigner-Seitz cell can be better predicted if one implements the 
Thomas-Fermi or Hartree-Fock theory \cite{Bonche,Barranco,Ogasawara,Togashi}.  
Even in such predictions, the result still depends on uncertainties in the 
adopted equation of state of asymmetric nuclear matter, especially the density 
dependence of the symmetry energy.

Experimentally, there can be two ways of studying properties of hot nuclear 
matter: heavy ion collisions at intermediate energy and quantum simulations 
with ultracold atoms.  Data for such heavy ion collisions have already given
some evidence for a liquid-gas phase transition of nuclear matter \cite{GSI}.
It is, however, important to note that the deduced temperature and density of
finite matter (primary fragments) created in the collisions have some 
uncertainties, while finite-size corrections due to the Coulomb and surface 
effects have to be taken into consideration to deduce the critical point of 
bulk nuclear matter from the empirical data for the excitation energy, mass,  
and temperature of the primary fragments.  Anyway, this is a direct way of
probing nuclear matter.  On the other hand, ultracold atoms provide us 
with an indirect way of probing nuclear matter.  This is mainly because 
low density neutron matter, which is dominated by $s$-wave scattering
with negative and large scattering length, is similar to trapped 
ultracold Fermi atoms near a Feshbach resonance \cite{Navon, Horikoshi}.
It is thus expected that various superfluid properties of neutron matter
such as the pairing gap, the BCS-BEC crossover, and vortices 
under rotation could be deduced from laboratory experiments, 
although the neutron-neutron effective range is fairly large
as compared with the case of atoms.  
Incidentally, effects of a nonzero effective 
range of the interaction and nonzero proton fraction on the 
equilibrium properties of matter relevant to neutron stars 
have been studied by bearing in mind the quantum 
simulation in cold atomic systems 
\cite{vanWyk2018,Tajima2019}. 
Moreover, one can add impurity
atoms to a system of ultracold Fermi atoms of a single species
\cite{Schirotzek_2009,Kohstall_2012,Scazza}.  In the presence of interspecies interactions, these 
impurity atoms manifest themselves as polarons.  The resultant 
atomic matter looks like cold nuclear matter of interest here. 
In relation to the present interest in the $\alpha$ particle as an impurity in cold neutron matter,  
there are cold atomic systems useful for the quantum simulation, e.g., 
a resonantly interacting atomic gas mixture of $^{161}$Dy and $^{40}$K \cite{Ravensbergen2020}, 
which has almost the same mass ratio with the $\alpha$-neutron system, 
a Fermi polaron interacting with medium via $p$-wave Feshbach resonance as well as background $s$-wave scattering 
\cite{Wille2008,Levinsen2012}, etc.

In the present study, under the motivation mentioned above, 
we evaluate in-medium modifications of an $\alpha$ particle that
interacts with surrounding neutrons, by calculating its quasiparticle properties: 
the interaction energy, residue, effective mass, dispersion relation, and decay width. 
To this end,  for simplicity, we employ an ideal situation 
where a single $\alpha$ particle is immersed in pure neutron matter 
that is in a normal state at zero temperature, rather than hot nuclear matter. 
Furthermore, we assume that the $\alpha$ particle is mobile with a small momentum
and that the neutron matter is dilute enough, 
so that we can employ the low-energy treatment, i.e.,  the $\alpha$ particle is point like, 
while the interaction between the $\alpha$ particle and 
a surrounding neutron is described only by the $s$-wave scattering length. 
More quantitative discussion on these approximations will be given at the end of Sec.~IV.

Note that such an $\alpha$ particle is not always stable.  In fact,
stability analysis of an $\alpha$ particle in neutron matter would require 
microscopic calculations, which will be addressed in the near future.
We also remark that even before the experimental realization of
trapped ultracold atoms, Kutschera and Wojcik \cite{Kutschera} used to 
consider a proton 
impurity in neutron matter by sticking with the original polaron 
picture based on an electron-phonon system, which is different from
the modern polaron picture based on a minority atom - majority atom 
system involving contact interactions.

We also note that $\alpha$-nucleon interactions in hot nuclear matter
were considered in terms of excluded-volume effects \cite{Lalit}
and the virial expansion \cite{Horowitz}.  Essentially, the 
$\alpha$-nucleon interactions considered in these approaches tend to 
be too repulsive and too attractive, respectively, when applied to
a cold system of interest here. This is because
in the former approach, no attraction is included, while in the
latter approach, the second virial coefficient is related to the
two-body phase shift dominated by the $p$-wave resonance.

It is interesting to note the possible relevance of the 
present polaron picture to $\alpha$ clustering, i.e., manifestation 
of $\alpha$ particle like configurations, in atomic heavy nuclei.  
Although there is no direct evidence for the presence of such
$\alpha$ clustering, it is expected that the four-nucleon 
correlation responsible for the $\alpha$ clustering plays an 
essential role in describing the surface region of 
various heavy nuclei 
\cite{Brink1,Brink2,Yang2020}.

In such a dilute neutron-rich situation, two 
minority particles (protons) may tend to form
an $\alpha$ particle by picking two neutrons out of
the medium.  Consequently, the system may look 
like pure neutron matter containing $\alpha$ particles 
as impurities.  This situation may help us describe the 
neutron skin structure of heavy nuclei. 
The energy and decay rate of such $\alpha$ particles, if
being known experimentally, could give us an opportunity of 
probing the bulk properties of the neutron medium.

This paper is organized as follows: 
In Sec.\ II we present the low-energy effective Hamiltonian for a single $\alpha$ particle 
embedded in neutron matter. 
In Sec.\ III we employ a variational method to obtain the energy of the $\alpha$ particle, 
which is equivalent to the self-energy calculation from the ladder type approximation. 
In Sec.\ IV we first employ empirical scattering data for setup of 
the parameters in the Hamiltonian and then discuss the numerical results for various quasiparticle properties 
as functions of neutron density, and also of the scattering length for general arguments.  
Section V is devoted to summary, physical consequences, and outlook.

\section{formulation}
We consider a single $\alpha$ particle that is assumed to be an 
inert point particle and to be immersed in normal neutron matter 
at zero temperature.   The Hamiltonian of this system is described by 
\beq
H(x)&=&\sum_{s} \int {\rm d}r^3 \psi_s^\dagger(r)\frac{-\nabla^2}{2m}\psi_s(r)+ 
\frac{1}{2}\sum_{s, t, s', t'} \iint {\rm d}r^3{\rm d}{r'}^3 \psi^\dagger_{t'}(r') \psi_{s'}^\dagger(r)V_{s't'st}(r-r') \psi_s(r)\psi_t(r')
%%%
\nn 
&&
-\frac{\nabla_x^2}{2M}
+g\sum_{s} \int_r \psi_s^\dagger(r)\psi_s(r) \delta(r-x)
%%%
\\
&=& 
\sum_{s,p} \frac{p^2}{2m} a_{s,p}^\dagger a_{s,p} + 
\frac{1}{2}\sum_{q,q',p, s, t, s', t'}   
a_{t',q'-p}^\dagger  a_{s',q+p}^\dagger  a_{s,q} a_{t,q'}\tilde{V}_{s't'st}(p) 
%%%
\nn 
&&
-\frac{\nabla_x^2}{2M} + g \sum_{p,q,s} a_{s,p}^\dagger a_{s,q} e^{-i(p-q)x},
\label{H(x)}
\eeq
where $m$ ($M$) is the mass of a neutron (an $\alpha$ particle), 
and we have used the first quantization 
for the single $\alpha$ particle in the coordinate representation by $x$, 
and expanded the neutron field operator as 
$\psi_s(r)=\frac{1}{\sqrt{V}}\sum_p e^{ipr} a_{s,p}$ 
with the canonical relation 
$\ltk a_{s,p}^\dagger, a_{t,q}\rtk=\delta_{p,q}\delta_{s,t}$, 
where the subscript $s$ and $p$ represent the neutron spin 
and momentum, respectively. 
It is noted that the bare coupling constant $g$ for the $\alpha$-neutron interaction is related to 
the scattering length $a$ via the Lippmann-Schwinger equation in the low-energy limit: 
\beq
g^{-1}&=& 
\frac{m_r}{2\pi \hbar^2 a} -\sum_{p} \frac{2m_r}{p^2},  
\label{LS1}
\eeq
where ${m_r}^{-1}=m^{-1}+M^{-1}$ is the reduced mass, 
and a momentum cutoff $\Lambda$ is assumed implicitly in the divergent integral $\sum_{p}$ as a regulator.
The cutoff $\Lambda$ corresponds to the effective range $r_0$ via $r_0\sim \Lambda^{-1}$, and can be sent to infinity 
after the renormalization of the bare coupling constant $g$ in terms of the scattering length $a$ in Eq.~(\ref{self1}).  
We use the natural units in which $\hbar=c=1$.

Now we implement a gauge transformation \cite{LLP1},  
\beq
S(x) &:=& e^{ix \hat{P}},  \ 
\mbox{ with } \hat{P} = \sum_{s,p} p a_{s,p}^\dagger a_{s,p}, 
\eeq
which sends a gas of neutrons to the comoving frame 
of the impurity $\alpha$ particle.  
By using 
$S(x) a_{s,p}S^{-1}(x) = a_{s,p} e^{-ipx}$, 
the Hamiltonian  (\ref{H(x)}) can be transformed to 
\beq
SH(x)S^{-1}
&=&  
\sum_{s,p} \frac{p^2}{2m} a_{s,p}^\dagger a_{s,p}+ 
\frac{1}{2}\sum_{q,q',p,s,t}  a_{t,q'-p}^\dagger  a_{s,q+p}^\dagger  a_{s,q} a_{t,q'}\tilde{V}_{s,t}(p) 
%%%
\nn 
&&
+\frac{\lk-i\nabla_x -\hat{P}\rk^2}{2M}+g \sum_{p,q,s} a_{s,p}^\dagger a_{s,q}, 
\eeq
which satisfies the commutation relation 
$\ldk SH(x)S^{-1} , -i\nabla_x \rdk=0$,
implying that after the transformation the momentum of the $\alpha$ particle represents the total momentum of the system. 
Therefore, we can replace the momentum operator by a $c$-number vector, i.e., 
$ -i\nabla_x \rightarrow P$ in the transformed Hamiltonian, 
\beq
SH(x)S^{-1}\ \  \rightarrow \ \  H_{\rm eff}(P) = 
\sum_{s,p} \varepsilon_{p} \,  a_{s,p}^\dagger a_{s,p} 
+\frac{\lk P-\hat{P}\rk^2}{2M}+g \sum_{p,q,s} a_{s,p}^\dagger a_{s,q}, 
\label{Heff1}
\eeq
where we have furthermore dropped the $\tilde{V}_{s,t}$ 
by assuming that the neutron self-interaction effects 
are incorporated in the single-particle energy
$\varepsilon_{p}=\frac{p^2}{2m^*}+U$.  
Here, 
$m^*$ is the effective mass in medium, 
but we assume that the in-medium modification is negligible 
and take the same notation, i.e., $m^*=m$ hereafter.
$U$ is the density dependent interaction energy per particle, 
which, e.g.,  can be deduced from a Thomas-Fermi approach 
 \cite{OI} but is not relevant for the present study.  
We will use the above Hamiltonian (\ref{Heff1}) 
in the following calculations. 

\section{single particle-hole pair approximation} 
To describe excitations accompanied by the impurity $\alpha$ particle,
we implement a variational method in which the variational state 
is spanned by a single particle-hole (p-h) pair 
excitation near the neutron Fermi surface \cite{Chevy1, Chevy2}:
\beq
|\Psi\rangle 
= F_0 |\psi_0 \rangle + \sum_{k>,p<, s} F_{k,p}^s a_{s,k}^\dagger a_{s,p} |\psi_0 \rangle,
\label{Vstate1} 
\eeq
where $|\psi_0\rangle$ is the state occupied by neutrons up to 
the Fermi momentum $k_F$, $k>$($p<$) denotes $|k|>k_F$ ($|p|<k_F$), 
and $F_0$ and $F_{k,p}^s$ are variational parameters. 
In fact, $F_{k,p}^s$ serves as the wave function of 
the p-h pair of the corresponding momentum and spin.
We remark in passing that even such a lowest-order form of
the variational state can well reproduce the empirical energy and 
effective mass of an impurity that repulsively interacts with medium 
fermions in the vicinity of the unitarity limit, i.e., $|a|\to\infty$ and 
zero effective range \cite{Scazza}.

The expectation value of the transformed Hamiltonian (\ref{Heff1}) 
with respect to the state (\ref{Vstate1}) gives 
\beq
\langle H_{\rm eff} (P)\rangle 
&=&  
\sum_{s,q} \varepsilon_{q} \langle  a_{s,q}^\dagger a_{s,q} \rangle 
+\frac{\left\langle \lk P-\hat{P} \rk^2\right\rangle }{2M}
+g\sum_{q,q',s} \langle a_{s, q}^\dagger a_{s, q'} \rangle,
\eeq
where $\langle\cdots\rangle = \langle \Psi|\cdots|\Psi\rangle$, 
and each term in the right side is given, respectively, by 
\beq
\sum_{q} \varepsilon_{q} \langle a_{s,q}^\dagger a_{s,q} \rangle 
&=&
\sum_{q<} \varepsilon_{q} \lk |F_0|^2+\sum_{k>,p<, t} |F_{k,p}^t|^2 \rk 
+
\sum_{k>,p<} \lk \varepsilon_{k}-\varepsilon_{p}\rk |F_{k,p}^s|^2, 
\\
%%%%%%%%%%%%%%
\left\langle \lk P-\hat{P} \rk^2\right\rangle 
&=&  P^2 |F_0|^2
+ \sum_{k>,p<,, s} |F_{k,p}^s|^2\ldk P^2-2 P\cdot (k-p)+(k-p)^2\rdk, 
\\
%%%%%%%%%%%%%
\mbox{ and } 
\sum_{q,q'} \langle a_{s,q}^\dagger a_{s,q'} \rangle 
&=&  \sum_{q<} \lk |F_0|^2+\sum_{k>,p<, t} |F_{k,p}^t|^2 \rk   
+ \sum_{k>,p<} \lk F_0F_{k,p}^{s*} + F_0^* F_{k,p}^s\rk 
\nn
&&
+ \sum_{q>,q'>,p<} F_{q,p}^{s*}F_{q',p}^s-\sum_{k>,q<,q'<} F_{k,q}^{s*}F_{k,q'}^s.  
\eeq

\subsection{Quasiparticle energy}
We impose the normalization condition by using a Lagrange multiplier 
$\mu$ that turns out to be the ground state energy $E_P$ 
of the system with momentum $P$. 
In fact,  $E_P$ represents the $\alpha$ particle dispersion relation 
up to the Fermi energy of the medium neutrons.
The variational condition $\delta \langle H_{\rm eff} -\mu \rangle=0$ 
leads to a set of eigenvalue equations, 
\footnote{
One can also obtain the same result from a time-dependent 
variational approach to the Dirac type effective action, i.e., 
$\delta\langle \Psi(t)| i\partial_t -H_{\rm eff} |\Psi(t) \rangle=0$, 
if one assumes $|\Psi(t) \rangle \sim e^{-i\mu t}$.  Note that
this $|\Psi(t) \rangle$ includes excited states, as well as
the ground state. }
\beq
\frac{P^2}{2M} F_0+\sum_{q<,s} g \lk F_0 +\sum_{k>} F_{k,q}^s\rk &=& \omega F_0, 
\label{eve1}
%%%%%%%%%%%%%%%
\\ 
\Omega_{k,p;P}^s F_{k,p}^s
+g \lk F_0 +\sum_{q'>} F_{q',p}^s-\sum_{q'<} F_{k,q'}^s\rk &=& \omega   F_{k,p}^s, 
\label{eve2}
\eeq
where $\omega=\mu-\sum_{q<,s}\varepsilon_{q}$, 
and 
\beq 
\Omega_{k,p;P}^s &:=& 
\varepsilon_{k} -\varepsilon_{p} 
+\frac{P^2-2 P\cdot (k-p)+(k-p)^2}{2M}+g \sum_{q<}. 
\eeq
From the Lippmann-Schwinger equation (\ref{LS1}), 
the bare coupling constant $g$ turns out to be vanishingly small 
as a negative power of the momentum cutoff $\Lambda$. 
In the renormalization procedure in terms of the scattering length \cite{Chevy1},
therefore, we will drop the terms 
$g\sum_{q'<} F_{k,q'}^s$ in Eq.~(\ref{eve2}) and 
$g\sum_{q<}$ in $\Omega_{k,p;P}^s$ as sub leading order.

Using the auxiliary field $\chi_p^s =  F_0 +\sum_{q'>} F_{q',p}^s$ 
in  Eqs.~(\ref{eve1})--(\ref{eve2}), 
we obtain the following equation: 
\beq
\omega&=&\frac{P^2}{2M} +\Sigma\lk \omega, P\rk, 
%%%%%%%%%%
\\ 
\mbox{with } 
\Sigma\lk \omega, P\rk
&=&
 \sum_{p<, s} 
\frac{1}{ \frac{m_r}{2\pi a} -\sum_{k>} \lk \frac{1}{\omega-\Omega_{k,p;P}^s}+\frac{2m_r}{k^2}\rk -\sum_{q<} \frac{2m_r}{q^2}},
\label{self1}
\eeq 
from which the eigenvalues of $\omega$ can be determined. 
In this equation, $\Sigma\lk \omega, P\rk$ can be interpreted as 
the self-energy obtained from the non-selfconsistent ladder 
approximation for the $\alpha$-neutron scattering amplitude \cite{Chevy2}. 
In the case of a repulsive interaction $a>0$ of interest here, 
the resulting positive energy state is characterized by
the outgoing scattering amplitude.   The real part of the corresponding 
quasiparticle energy $E_P$ can thus be obtained from the spectral 
peak as 
\beq
E_P &=&\frac{P^2}{2M} +{\rm Re}\Sigma\lk E_P+i0, P\rk, 
\label{real1} 
\eeq
where the analytic continuation to the upper half plane 
$\omega\rightarrow \omega+i0$ has been made; it traces back to the 
(retarded) propagator of the $\alpha$ particle that
undergoes multiple scattering.

\subsection{Quasiparticle residue, width, and effective mass} 
Validity of the quasiparticle picture for the $\alpha$ particle 
requires a finite residue and a relatively small width 
(long lifetime) compared with the real part of 
quasiparticle energy, both of which imply that the propagator 
of the $\alpha$ particle near its pole behaves as
\beq
G^R(\omega,P) &=& \frac{1}{\omega+i0 -\frac{P^2}{2M}-\Sigma(\omega+i0,P)} 
\sim \frac{Z_P}{\omega -E_P+i \Gamma_P},
\eeq
where the width is given approximately by the imaginary part, 
$\Gamma_P=-Z_P {\rm Im} \Sigma\lk E_P+i0, P\rk$, 
and the quasiparticle residue is defined by 
\beq
Z_P=
\ldk 
1
-\left. {\rm Re} \frac{\partial \Sigma(\omega+i0,P)}{\partial \omega}\right|_{\omega=E_P} 
\rdk^{-1}. 
\eeq
\footnote{
Unlike unitary cold atoms there is no bound state 
between an $\alpha$ particle and a neutron.  In fact,
a resonant $^5$He (3/2-), if appearing in vacuum, would be very 
unstable to $p$-wave dissociation into an $\alpha$ particle 
and a neutron, while in the $s$-wave channel there is no positive 
scattering length state that can be obtained continuously 
from the negative scattering length state (attractive branch). 
In the absence of decay to the attractive branch, 
therefore, the present positive energy state is not an excited one. 
}
The effective mass $M^*$, which characterizes the mobility of 
the $\alpha$ particle in the medium, is defined in the momentum 
expansion around the pole as 
\beq
M^*:=\lk \left. \frac{{\rm d}^2 E_P}{{\rm d}P^2}\right|_{P=0}\rk^{-1}
=\frac{M}{Z}
\ldk 
1+M\left. \frac{\partial^2 {\rm Re} \Sigma(\omega+i0,P)}{\partial P^2}\right|_{\omega=E,P=0} 
\rdk^{-1}, 
\eeq 
where $E=E_{P=0}$ and $Z=Z_{P=0}$. 
As will be shown by numerical results below, the quasiparticle picture is 
indeed valid for the parameter region that corresponds to the 
low density and large isospin asymmetric situation considered in this study.
The quasiparticle dispersion of the $\alpha$ particle can thus be well 
approximated by 
\beq 
E_P \simeq E + \frac{P^2}{2M^*} 
\label{apprdis1}
\eeq  
near the low energy limit.  

\section{Numerical results and discussion} 
Before exhibiting numerical results for the quantities described
in the previous section, we give typical reference values of 
the scattering length and the neutron density as
\beq
a_{\rm ref} &=& 2.64 \ {\rm fm}, 
\label{scat1} \\
\rho_{\rm ref} &=& 0.01 \, \rho_0, \quad \rightarrow \quad k_{F {\rm ref}} = 0.36 \ {\rm fm}^{-1}, 
\eeq 
where $\rho_0=0.16$ fm$^{-3}$ is the normal nuclear density. 
These values lead to the dimensionless coupling parameter  
$a_{\rm ref} k_{F {\rm ref}} =0.95 \simeq 1$, 
which marginally correspond to the strong coupling regime.  
The value of $\rho_{\rm ref}$ is supposed to be of order 
the typical density above which neutrinos are trapped in 
supernova cores \cite{Bethe}, but lower than the typical  
density in the neutron skin of heavy nuclei \cite{Typel2}. 
The value of the scattering length has been determined 
from the $\alpha$-neutron potential \cite{Kanada} that 
reproduces the experimental phase shift for 
low-energy $\alpha$-neutron scattering cross section.
In the course of this determination, the effective range  
has been simultaneously obtained as $r_0 = 1.43$ fm. 
This value gives $r_0 k_{F {\rm ref}} = 0.51 <1$, which is
not so small but still implies the validity of the present 
low-energy treatment of the $\alpha$-neutron interaction
in terms of the zero-range potential characterized by 
the scattering length (\ref{LS1}) alone.
We remark that the values of $a_{\rm ref}$ and $r_0$ obtained 
here are consistent with the earlier results \cite{Arndt1974,Kamo1}. 
Finally, we set the bare mass ratio of the $\alpha$ particle as 
\beq
\frac{M}{m} &=& 4. 
\eeq

\begin{figure}[h]
  \begin{center}
    \begin{tabular}{cc}
 \resizebox{90mm}{!}{\includegraphics{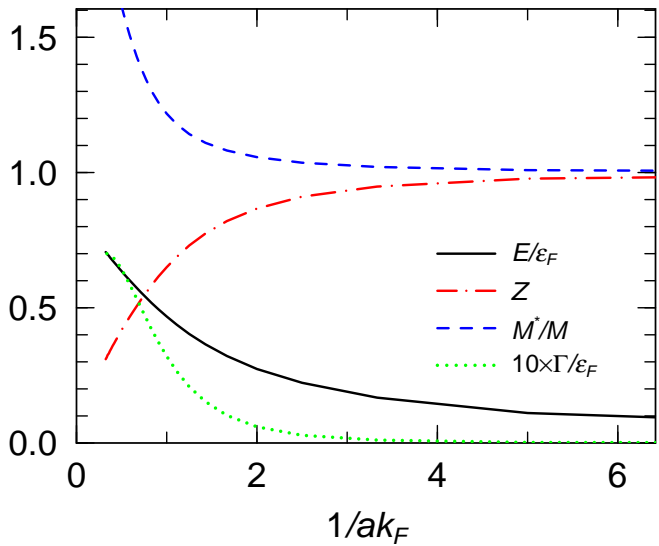}} %& 
 %\resizebox{90mm}{!}{\includegraphics{alphaEp_data1_Fig.eps}} %\\
% \resizebox{50mm}{!}{\includegraphics{nemo2.eps}} & 
% \resizebox{50mm}{!}{\includegraphics{nemo1.eps}} \\ 
    \end{tabular}
    \caption{Energy, residue, and mass ratio 
of an $\alpha$ particle calculated
as functions of $1/ak_F$ at $P=0$.}
    \label{fig1}
  \end{center}
\end{figure}

For these reference values of the parameters given above, 
the numerical results for quasiparticle properties of the $\alpha$ particle at $P=0$ are given as follows:
\beq
E_{\rm ref} &=&  0.467 \, \varepsilon_{F {\rm ref}} =1.26 \mbox{ MeV}, 
\label{intene1}\\ 
Z_{\rm ref} &=& 0.650, \\
M^*_{\rm ref} &=&1.217 \, M, \\
\mbox{and } \ 
\Gamma_{\rm ref} &=&  0.032 \, \varepsilon_{F {\rm ref}}=0.086 \mbox{ MeV},
\eeq
where $\varepsilon_{F {\rm ref}}=\frac{(\hbar c)^2 k_{F {\rm ref}}^2}{2m c^2}=2.69$ MeV. 
The interaction energy of the $\alpha$ particle at rest, $E_{\rm ref}$,
is positive, a feature that reflects the positive scattering length
corresponding to repulsion, and is much larger than the decay width.  
As can be interpreted from the diagrammatic representation of the present 
self-energy, the decay width comes solely from the process in which  
the $\alpha$ particle, a quasiparticle dressed by a cloud of p-h bubbles, 
decays into a bare $\alpha$ particle and a neutron in the $s$-wave 
scattering state. 
In addition, the quasiparticle residue obtained here is fairly 
close to unity.  All these results 
support the quasiparticle picture presumed in this study. 
At the reference neutron density 
the effective mass increases by about $20 \%$ from its vacuum value.
Such increase seems like a general feature of various 
impurity-medium combinations, irrespective of whether the impurity-medium 
interaction is repulsive or attractive. 
While the scattering length (\ref{scat1}) employed here is an empirical one,  
for more general discussion, we extend our calculations and present 
in Fig.~\ref{fig1} the results as functions of the dimensionless 
coupling.
%%%%%%%%%%%%%%%%%%%%%%%%%%%%%%%%%%%%%%%%%%%%%%%%%%%%%%%%%%%%%
%%%%%%%%%%%%%%%%%%%%%%%%%%%%%%%%%%%%%%%%%%%%%%%%%%%%%%%%%%%%%%
For weak coupling %$ak_F\ll 1$, 
the quasiparticle picture works obviously.
It is also found from the figure that 
while even in the strong coupling regime $1/ak_F\ll1$, 
the relation $E \gg \Gamma$ still holds, in such a regime 
the quasiparticle is less identifiable 
due to the smallness of the residue.

\begin{figure}[h]
  \begin{center}
    \begin{tabular}{cc}
% \resizebox{90mm}{!}{\includegraphics{alphaEnergyZRatio_data1_Fig.eps}} & 
 \resizebox{90mm}{!}{\includegraphics{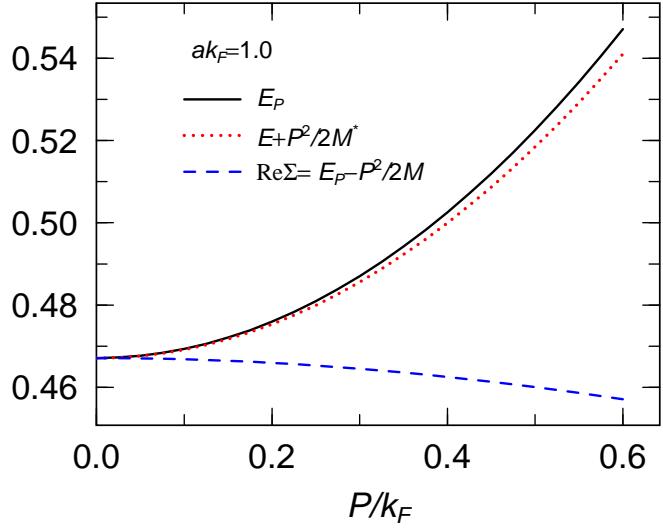}} %\\
% \resizebox{50mm}{!}{\includegraphics{nemo2.eps}} & 
% \resizebox{50mm}{!}{\includegraphics{nemo1.eps}} \\ 
    \end{tabular}
    \caption{Dispersion relations of an $\alpha$ particle calculated 
in units of 
$\varepsilon_F$ at $ak_F=1.0 \simeq a_{\rm ref}k_{F{\rm ref}}$.}
    \label{fig2}
  \end{center}
\end{figure}

Now we show in Fig.~\ref{fig2} the full dispersion relation $E_P$ obtained from 
Eq.~(\ref{real1}) with the reference values of the parameters, 
together with the approximate one (\ref{apprdis1}) expressed by  
the effective mass. 
%%%%%%%%%%%%%%%%%%%%%%%%%%%%%%%%%%%%%%%%%%%%%%%%%%%%%%%%%%%%%
%%%%%%%%%%%%%%%%%%%%%%%%%%%%%%%%%%%%%%%%%%%%%%%%%%%%%%%%%%%%%%
We observe that the appreciable $P$ dependence of the self-energy 
${\rm Re}\Sigma$, also plotted in the figure, accounts for 
the deviation between the full and the approximate dispersion 
relations for finite $P$. 
The deviation is nevertheless small enough that 
the approximate one seems to work for a relatively wide range 
of the momentum.

Finally we show how the quasiparticle properties depend on 
the neutron density. 
%%%%%%%%%%%%%%%%%%%%%%%%%%%%%%%%%%%%%%%%%%%%%%%%%%%%%%%%%%%%%
\begin{figure}[h]
  \begin{center}
    \begin{tabular}{cc}
% \resizebox{90mm}{!}{\includegraphics{alphaEnergyZRatio_data1_Fig.eps}} & 
 \resizebox{90mm}{!}{\includegraphics{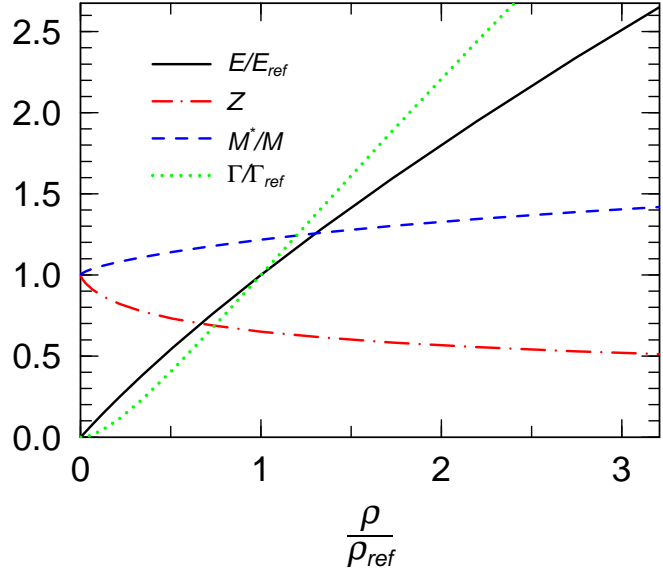}} %\\
% \resizebox{50mm}{!}{\includegraphics{nemo2.eps}} & 
% \resizebox{50mm}{!}{\includegraphics{nemo1.eps}} \\ 
    \end{tabular}
    \caption{Neutron density dependence of the
energy, residue, effective mass, and width of an $\alpha$ particle
calculated at $ak_{F{\rm ref}}=1.0$. }
    \label{fig3}
  \end{center}
\end{figure}
%%%%%%%%%%%%%%%%%%%%%%%%%%%%%%%%%%%%%%%%%%%%%%%%%%%%%%%%%%%%%%
Since the increase in the density extends the phase space 
available for the p-h fluctuations, 
the interaction energy and the decay width increase with density,
as can be seen in Fig.~\ref{fig3}.  On the other hand,   
the relation $E\gg \Gamma$ is kept, and eventually 
at $\rho=4.2\rho_{\rm ref}=0.0067$ fm$^{-3}$, the kinetic energy of the background 
neutron gas reaches 
$\varepsilon_F=\varepsilon_{F{\rm ref}} \lk \frac{\rho}{\rho_{\rm ref}} \rk^{2/3} =7$ MeV, 
which is added to the kinetic energy of the constituent neutrons inside the $\alpha$ particle, 
and thus cancels the binding energy of the $\alpha$ particle per nucleon $\sim7$ MeV.
At this stage, the stability of the $\alpha$ particle itself becomes doubtful. 
More realistically, the neutron separation energy of $^4$He in vacuum, which 
amounts to 20.6 MeV, helps us to give a better estimate of the critical neutron 
density at which the $\alpha$ particle is no longer bound.  At the neutron density of
about $21 \rho_{\rm ref}\simeq 0.034$ fm$^{-3}$, the neutron Fermi energy 
reaches 20.6 MeV.  Interestingly, 
a microscopic calculation for the energy of the four-nucleon system in 
symmetric nuclear matter \cite{Roepke2014} has shown
by comparison with the free four-nucleon energy threshold in the medium
that the $\alpha$-like cluster dissociates at $\rho = 0.03$ fm$^{-3}$ 
due to Pauli blocking effects at zero temperature, 
which is close to the above neutron density.

In what follows we discuss the negligibility of higher partial-wave contributions 
than the $s$ wave. 
In general the scattering $T$ matrix and its partial-wave decomposition \cite{Taylor1974} 
can be obtained from the Lippmann-Schwinger equation as 
\beq
T({\bf k},{\bf k}')=\sum_{l=0,1,2,\cdots} (2l+1) T_l(k) P_l(\hat{k}\cdot \hat{k}'), 
\label{Tmat1}
\eeq
with $P_l(x)$ being the Legendre polynomials, 
and the partial-wave matrix can be expanded, 
at low scattering energy $E_{\rm rel}=k^2/2m_r$ and for finite range two-body potentials, as 
\beq
T_l(k) = \frac{2\pi}{m_r} \, \frac{k^{2l}}{a_l^{-1} - r_l k^2/2 + {\mathcal{O}}(k^4) + ik^{2l+1}}, 
\label{Tmat2}
\eeq
where $a_l$ and $r_l$ correspond respectively to the generalized 
scattering length and effective range just like the case of the $s$ wave; 
$a_0$ is identical with $a$. 
The quasiparticle poles for possible bound/resonance states at low energies 
can be obtained in terms of these parameters. 
Now let us give a ballpark estimate of the density region in which 
the $s$ wave dominates in the scattering process: 
Comparing the $s$- and $p$-wave matrices with the empirical scattering length (volume) at the 
neutron Fermi momentum, 
the condition $|T_0| > |T_1|$ reduces to 
\beq
a_0 > |a_1| (m_r/m)^2 k_F^2\ \, \rightarrow \ \, k_F < 0.25\, {\rm fm}^{-1} \ \, \rightarrow \ \, \rho < 0.003 \rho_0, 
\eeq 
which is compatible with the validity condition for the Taylor expansion in (\ref{Tmat2}), i.e., 
$2|a_0|^{-1} > |r_0| (m_r/m)^2 k_F^2$ and $2|a_1|^{-1} > |r_1| (m_r/m)^2 k_F^2$.
Here we have used $a_0=2.64$ ${\rm fm}$ ($a_1=a_{P_{3/2}}=-67.1$ ${\rm fm}^3$) 
and $r_0=1.43$ ${\rm fm}$  ($r_1=r_{P_{3/2}}=-0.87$ ${\rm fm}^{-1}$) for the $s$-($p$-) wave scattering amplitude, 
obtained from the phenomenological $\alpha$-$n$ potential and scattering data 
in vacuum \cite{Kanada}. 
Furthermore, the $p$-wave  ($P_{3/2}$) resonance in the $\alpha$-$n$ system appears 
at the center-of-mass neutron kinetic energy $\sim 0.9$ MeV \cite{Horowitz,Arndt1974},
which amounts to the neutron Fermi momentum $k_F\sim 0.26$ ${\rm fm}^{-1}$ 
when the $\alpha$ particle is at rest. 
At densities higher than the one corresponding to $k_F\sim 0.26$ fm$^{-1}$, 
the $P_{3/2}$ resonance ($^5$He) may survive strong decay long enough 
to constitute a fraction of the matter 
under the influence of Pauli blocking effect \cite{Roepke2020}, 
although this is a scenario valid if and only if the $\alpha$ particle itself is 
bound in neutron matter and also the resonance energy is the same as in vacuum.
We remark in passing that if a similar argument can be applied 
to the temperature direction, 
the present model is valid at temperatures below both about $1$ MeV 
and the neutron Fermi temperature.

The above estimates indicate that the present model, which neglects 
higher waves, is safely applicable only at very low neutron densities $\rho < 0.003 \rho_0$, 
although this bound is obtained from the extrapolation 
from the zero density limit using the two-body scattering data in vacuum. 
Nevertheless, we stress that the quasiparticle picture of the $\alpha$ particle, 
which is obtained from the $s$-wave pole structure in this study,  
seems hardly disturbed by the other higher-wave contributions, 
as seen from the independence of different $T_l$'s in (\ref{Tmat1})
that is valid even at finite densities. 
For realistic description of the system at even higher 
densities, of course, 
we have to pursue the quasiparticle poles from the $p$- and higher-wave matrices $T_{l>0}$ as well as the $s$-wave one, 
and compare quasiparticle energies and strengths from these poles to figure out which one dominates at given density,  
or we can calculate the $T$ matrix directly using an empirical two-body potential without the partial-wave decomposition.
This is, however, out of the present scope.

We finally remark that in a very wide density regime below the normal nuclear density, 
zero-temperature supernova matter at proton fraction of order $0.3$ can satisfy 
the condition that the Fermi momentum of a neutron gas outside nuclei is below 
$0.25$ fm$^{-1}$ (see Fig. 4 in Ref.\ \cite{bastian15}).  We thus believe that
the present model is fairly reasonable from a phenomenological point of view.

\section{Summary and outlook} 
In this study we elucidate the quasiparticle properties of a 
single $\alpha$ particle immersed in a cold dilute neutron gas 
by evaluating the self-energy from the variational treatment 
equivalent to a non self-consistent ladder approximation 
that incorporates an empirical value of the $\alpha$-neutron 
low-energy $s$-wave scattering length (\ref{scat1}). 
Our result shows that adding a single $\alpha$ particle 
into a dilute neutron gas costs at least the interaction energy 
(\ref{intene1}) that we calculated for the $\alpha$ particle at rest.   
Note that such interaction energy reads 
$E=0.467 \varepsilon_F = \frac{2.23}{m} \rho^{2/3}$. 
This could be useful to deduce the fraction of $\alpha$ particles
in supernova matter.  It is also interesting to note that 
the effective mass calculated with the same parameter set leads to 
an approximate dispersion relation of the $\alpha$ particle,  
$E_p=E+P^2/2M^*$, with the density-dependent interaction energy $E$ 
given above.  Applying the dispersion relation to the Bose distribution 
for a dilute $\alpha$ gas in neutron matter, we can estimate the transition 
temperature for possible Bose-Einstein condensation \cite{Wu} to be 
$T_{\rm bec}=\frac{2\pi \hbar^2}{M^*}\lk \frac{\rho_\alpha}{\zeta(3)}\rk^{2/3}$ 
MeV, which gives $T_{\rm bec}=0.14$ MeV at $\rho_{\alpha}=\rho_{\rm ref}/10$. 
This might have some relevance to $\alpha$ clustering in the surface 
of heavy nuclei. 

%Outlook: 
%1) single $\alpha$: its wave function and  stability: 
In the present work the $\alpha$ particle in neutron matter is treated as 
an inert point particle, i.e., the inner structure is neglected 
from  the low-energy point of view.  To go beyond it, 
the $\alpha$ particle has to be regarded as a cluster of four
nucleons.  Then, diagrammatically needed is to take into account 
neutron exchange in and out of the cluster in scattering processes
as well as deuteron-like pair correlations, which might eventually 
break up the $\alpha$ particle at sufficiently high densities 
\cite{Roepke1,Beyer1,Typel1}.  
For elaborate studies to clarify 
the stability of such $\alpha$ clustering in neutron matter,  
the wave function of the system is needed in terms of interacting 
nucleons with phenomenological potentials 
\cite{OI,Roepke2014}.

%2) Many-body effects of $\alpha$: 
As discussed in the previous section, 
the present study is restricted to the dilute limit of the $\alpha$ density 
and to the cold, isospin asymmetric limit of the nuclear medium. 
Next it is interesting to consider the system of two $\alpha$ particles
in neutron matter.  The interaction between the $\alpha$ particles,
which resembles that between two repulsive Fermi polarons in the context
of ultracold atoms except for the Coulomb repulsion, has some possible 
relevance to neutrino scattering off light clusters in supernova matter. 
To make better estimates, it would be necessary to raise both the proton 
fraction and temperature \cite{Tajima_2018} in the nuclear medium and also consider screening 
corrections to the $\alpha$-$\alpha$ Coulomb repulsion \cite{Horiuchi1}.

\acknowledgments

We are grateful to T.~Hatsuda, P.~Naidon, H.~Tajima, T.~Uesaka, J.~Zenihiro, 
K.~Nakazato, and H.~Togashi for useful 
discussion.  This work was in part supported by Grants-in-Aid for Scientific 
Research through Grant Nos.\ 17K05445, 18K03635, 18H01211, 18H04569, 18H05406, 
and 19H05140, provided by JSPS. 
Author E.~N. acknowledges the hospitality of the 
Institut f\"{u}r Theoretische Physik, Goethe Universit\"{a}t Frankfurt, 
where this work was completed.
Author W.~H. acknowledges the Collaborative Research Program 2020, Information Initiative Center, Hokkaido University.

%E.~N. and  are supported by Grants-in-Aid for 
%through Grants No.~17K05445,and, respectively,  

%%%%%%%%%%%%%%%%%%%%%%%%
%\bibliography{Bosepolaronintrap_refs-v1}
%%%%%%%%%%%%%%%%%%%%%%%%%%%%%%%%%%%%%%%%%%%%%%%%%%%%%%%%%%%

\appendix
\section{Dimensionless expression for the self-energy} 
It is convenient to 
rewrite the self-energy in terms of dimensionless quantities (symbols with tilde) 
as well as $k_F$ and $\varepsilon_F$ as  
\beq
\Sigma(\omega+i0,P) &=&
\sum_{p<} \frac{1}{g_r^{-1}-G(\omega+i0, P; p)-\sum_{q<}\frac{2m_r}{q^2}}
\nn
&=&
(2\pi)^2 \varepsilon_F \sum_{\rho<} \frac{\tilde{g}^{-1}-{\rm Re}\tilde{G}+i{\rm Im}\tilde{G}}
{\lk \tilde{g}^{-1}-{\rm Re}\tilde{G}\rk^2+{\rm Im}^2\tilde{G}}, 
\label{appSigma1}
\eeq
with 
\beq
G(\omega, P; p) &\equiv&
\sum_{k>} 
\ltk 
\ldk \omega-\varepsilon_k+\varepsilon_p-\frac{P^2-2P\cdot\lk k-p\rk +\lk k-p\rk^2}{2M}\rdk^{-1}
+\ldk \frac{k^2}{2m_r}\rdk^{-1}
\rtk. 
\nn 
\eeq 
Here we have factored out the dimensional coefficients  as 
$G = \frac{k_F^3}{(2\pi)^2\varepsilon_F}\tilde{G}$ and 
$g_r^{-1}=\frac{k_F^3}{(2\pi)^2\varepsilon_F}\tilde{g}^{-1}$, 
and expressed the rest parts as  
\beq
%%%%%%%%%%%%%%%
\nn 
\tilde{G}(\omega, P; p-P) &=&
\int_{1}^{\infty} {\rm d}\kappa  \kappa^2
\int_{-1}^{1} {\rm d}x\, 
\ltk
\ldk 
\mE- \kappa^2+\bar{\rho}^2-\frac{ \kappa^2+\rho^2-2\kappa\rho x}{R}
+i 0\rdk^{-1} 
+
\ldk 
\kappa^2 (R^{-1}+1)
\rdk^{-1}
\rtk, 
\nn 
\label{int1}
\\ 
%%%%%%%%%%%% 
\mbox{and } \quad && \nonumber \\
\tilde{g}^{-1}
&=& 2\pi^2 \frac{R}{R+1} \lk \frac{1}{2\pi a k_F}-\sum_{\rho<} \frac{2}{\rho^2}\rk, 
%\nonumber
\eeq
%%%%%%%%%%%%%%%%%%%%
where $\bar{\rho}\equiv \sqrt{(p-P)^2}/k_F$, and 
we have introduced variables: ${\mathcal E} =\omega/\varepsilon_F$, 
${\mathcal P} =P/k_F$, $R=M/m$, 
$|p|/k_F=\rho$, and $|k|/k_F=\kappa$, 
and shifted the momentum temporarily, $p \rightarrow p-P$, for convenience.   
Performing the $x$ integration in Eq.\ (\ref{int1}) first, we obtain  
\beq
&&\int_{-1}^1 {\rm d}x
\ldk 
\mE-\frac{R \kappa^2-R\bar{\rho}^2+\kappa^2+\rho^2-2\kappa\rho x}{R}
+i0\rdk^{-1} 
%%%%%%%%
\nn 
&=&
\frac{R}{2\kappa\rho}
\ldk 
\ln\left|
\frac
{(\kappa-\kappa_-)(\kappa-\kappa_+)}
{(\kappa+\kappa_-)(\kappa+\kappa_+)} 
\right| 
-i\pi 
\theta\lk 1-\left| \frac{R\mE-R \kappa^2+R\bar{\rho}^2-\kappa^2-\rho^2}{2\kappa\rho}\right|\rk 
\rdk, 
\eeq
where $\theta(x)$ is the Heaviside function, and 
\beq
\kappa_\pm &\equiv&
\frac{\rho}{R+1}
\pm \frac{1}{R+1}\sqrt{R(R+1)\bar{\rho}^2-R\rho^2+\mE R(R+1)}.  
\eeq
Note that $\kappa_+ \ge 0$ and  $\kappa_- \le 0$ for $R>1$ and $\mE>0$. 
Then, the real and imaginary parts read respectively as  
\beq
{\rm Re}\tilde{G}(\omega, P; p-P)
&=&
\int_{1}^{\infty} {\rm d}\kappa 
\ldk
\frac{R\kappa}{2\rho} 
\ln
\left| 
\frac
{(\kappa-\kappa_-)(\kappa-\kappa_+)}
{(\kappa+\kappa_-)(\kappa+\kappa_+)}
\right| 
+
2\frac{R}{R+1} 
\rdk
%%%%%%%%%
\nn
&=&
\frac{R}{4\rho}
\ldk
\lk1- \kappa_+^2\rk 
\ln\left|\frac{1+\kappa_+}{1-\kappa_+}\right|
+
\lk1- \kappa_-^2\rk 
\ln \left|\frac{1+\kappa_-}{1-\kappa_-}\right|
\rdk
-
\frac{R}{R+1}
\eeq
and 
\beq
{\rm Im}\tilde{G}(\omega, P; p-P) 
&=&
-\pi \int_{1}^{\infty} {\rm d}\kappa 
\frac{R\kappa}{2\rho}
\theta\lk 1-\left| \frac{R\mE-R \kappa^2+R\bar{\rho}^2-\kappa^2-\rho^2}{2\kappa\rho}\right|\rk 
%%%%%%
\nn 
&=&
-\frac{R\pi}{4\rho} 
\theta\lk \kappa_+ - 1\rk 
\ldk 
\kappa_+^2  
-\kappa_-^2 \theta\lk -\kappa_- - 1\rk 
-\theta\lk 1 + \kappa_- \rk 
\rdk.  
\eeq
The $p$ integration in (\ref{appSigma1}) can be done numerically 
after shifting back $p-P \rightarrow p$. 
\end{document}